\documentclass[aps,pre,epsfig,floats,twocolumn,superscriptaddress,amssymb,amsmath,floatfix,showpacs]{revtex4}
\usepackage{graphicx}
\usepackage{bm}  % bold math
\usepackage{amssymb}
\usepackage{color}
\usepackage{amscd}

\begin{document}
\title{Rectified motion in an asymmetric channel: the role of hydrodynamic interactions with walls}

\author{Behzad Golshaei}
\affiliation{Department of Physics, Institute for Advanced Studies
in Basic Sciences (IASBS), Zanjan 45137-66731, Iran}
\author{Ali Najafi}
\affiliation{
Physics Department, University of Zanjan, Zanjan 45371-38791, Iran}
\email{najafi@znu.ac.ir}

\date{\today}

\begin{abstract}
Dynamics of a  Brownian particle in an asymmetric micro-channel that is subjected to an external oscillating 
force is  numerically analyzed. 
In addition to the elastic collisions with the walls that are kind of short range interactions, the long range hydrodynamic 
influences of the walls have been considered. We demonstrate how the geometrical parameters of the channel change  the rectified current of the particle. 
As a result of numerical calculations, we show that 
long range hydrodynamic interactions with walls, decrease the efficiency of the Brownian ratchet.    
\end{abstract}

%         {05.40.-a}{Fluctuation phenomena, random processes, noise, and Brownian motion} 
%         {05.60.-k}{Transport processes} 
%         {47.15.G-}{Low-Reynolds-number (creeping) flows}

\maketitle

\section{Introduction}
For a Brownian particle that is in thermal equilibrium, second low of thermodynamics does not allow to achieve a rectified 
motion even when the particle is moving in an asymmetric periodic potential. 
In addition to symmetry breaking with an externally applied asymmetric periodic potential, the 
fluctuations need to obey an out of equilibrium statistics  to achieve a rectified motion at the microscopic world.  
The original idea of the rectification of 
random motions at microscopic scales dates back to the work by M. V. Smoluchovski \cite{smolo} where he discussed the issue of extracting 
useful work from fluctuations. R. P. Feynman has illustrated this idea by a very intuitive  gedanken 
experiment composed of ratchet and pawl \cite{feynman}. 
Very recently, a group of experimentalists have shown how the ratchet idea can be tested at macroscopic world by 
using a granular gas. In their  experiment, a  plate vibrating vertically provides a non equilibrium gas of granular particles. Collision of  these
granular particles with four vanes of an asymmetric rotary part, enforce it to rotate. The overall motion of the 
vanes is shown to be a rectified rotation in a direction preferred by the asymmetry of the 
vanes \cite{granular}.  

In addition to  the above fundamental interests on the physics of Brownian ratchet, it  also provides a basic physical mechanism for  
describing the dynamics of most  biophysical molecular motors \cite{ajdari}. 
\begin{figure}
\centering
\resizebox{0.45\textwidth}{!}{%
  \includegraphics{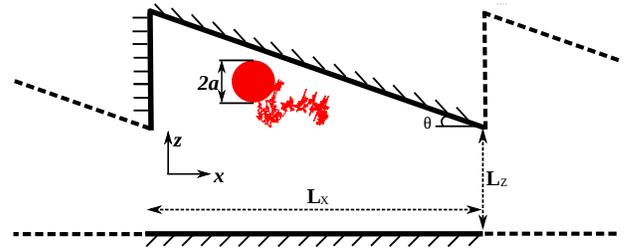}}
\caption{\small A  Brownian particle with radius $a$ moves in a two dimensional rocked Ratchet micro-channel with period $L_{x}$. 
The input and out of the channel have same width given by $L_{z}$. Angle $\theta$, measures the  asymmetry of channel. 
A typical trajectory of the Brownian particle, moving near the wall is shown.}
\label{fig1}
 \end{figure}
Transport of colloidal particles in channels with the size of micrometer is another related area. Recent technological advances allow researchers  
to design and fabricate 
devices to guide  particles on 
micro and nano-channels \cite{experimantal rocked ratchet,driftratchet}. 
Flow control and separation of particles in such channels are the main experimental interests in this field 
\cite{reimann2002brownian,particlespration}.
An important category of micro-fluidic devices is so-called rocked ratchet, where an  applied oscillating  
force drives a net drift velocity for particles fluctuating 
in an asymmetric potential imposed by the walls of channel . The physical mechanism behind such systems has been considered in details \cite{hanggi,martis}.
A very comprehensive review of the related works is presented   in an article by P. Hanggi \cite{review}.

Hydrodynamic interaction between particles, at intermediate and high volume fraction of particles is proved to have prominent effects on the efficiency 
of Brownian ratchets \cite{holger}. In addition to interaction between particles, the 
long range hydrodynamic interaction with confining walls is proved to have essential effects on the motion of either passive colloidal particles 
\cite{pralle,squires} or active systems \cite{roughwall,suspensionofconfined,3S2W}.
It is the main goal of this article to address how the interaction with walls will influence the functionality of a Brownian ratchet. 
Usually this kind of long range interactions are neglected in the theoretical and numerical investigations \cite{hnngiprl}. 
In this paper, we focus on hydrodynamic interactions of particles with the walls of channel.  To answer this question, 
we start by a  perturbation based theory that 
considers the hydrodynamic effects for a spherical particle moving near confining walls.  Then letting the particle to fluctuate, 
we numerically simulate the Brownian dynamics of such a colloidal particle moving in a medium confined with the walls of an asymmetric channel. 

The structure of this article is as follows: In section 2, we define the model and present the details of   approximations for the 
hydrodynamic interactions. In section 3, we present the details of numerical scheme that we have used to simulate the Brownian dynamics. 
The results are 
presented in section 4 and finally we discuss in section 5.

\section{Model and its parameters}
A two dimensional rocked ratchet channel is used to study the transport of Brownian particles. As depicted in fig.\ref{fig1}, the channel is 
characterized by its periodic length $L_x$, input and output opening sizes $L_z$  and asymmetry angle $\theta$. For $\theta=0$, 
the channel is symmetric and we do not expect any rectified motion for this case. 
Moreover we assume that the colloidal particle has  mass $m$ and its radius is given by $a$. In a reference frame that is shown in the figure, 
the position vector of the particle is 
given by: ${\bf r}=(x,z)$, where  ${\hat {\bf x}}$  points along the axis of channel. 
The following  stochastic Langevin differential equation  describes the dynamics of a colloidal particle moving in this channel:
\begin{equation}
m{\ddot {\bf r}}={\bf G}\cdot{\dot {\bf r}}+{\bf F}_{hc}+{\bf F}_e(t)+{\bf \Gamma}(t),
\label{langevin}
\end{equation}
where ${\bf G}$ stands for the hydrodynamic friction tensor and ${\bf \Gamma}(t)$ shows the random forces due to thermal fluctuations. 
The effects of the walls of channel are given by a  hard core force 
${\bf F}_{hc}$. As a result of this very short range potential, the  dynamics of the particle obeys the rules of  elastic collision 
at the boundaries. Conservation of energy and momentum provide relations for the state of the system before and after each collision.  
To provide an out of equilibrium condition for the particle, we apply an  external force along the 
${\hat {\bf x}}$ direction that is spatially uniform but period in time:  
\begin{equation}
{\bf F}_e(t)=f_0\sin(\omega t){\hat {\bf x}},
\end{equation}
where $f_0$ and $\omega$ show the amplitude and frequency of the external force, respectively. In addition to the above deterministic forces, the 
dynamics of the particle is influenced by time dependent random noise ${\bf \Gamma}(t)$ results from thermal fluctuations. 
The average and correlation of this stochastic force is given by:
\begin{equation}
\langle{\bf \Gamma}(t)\rangle=0,~~~~~\langle{\bf \Gamma}(t){\bf \Gamma}(t')\rangle=2k_BT{\bf G}\delta(t-t'),
\end{equation}
where $k_BT$ is the thermal energy. 

What we are interested in this article, is the effects of long range hydrodynamic interaction with the walls of the channel. 
All of  information related to the hydrodynamic interactions with the walls are encoded in the friction tensor ${\bf G}$. 
In addition to the size of colloid, this 
friction tensor depends on   distance between the particle and the walls. This friction coefficient, should in principle be 
derived from solution to Stokes equation, the equation that governs the dynamics of fluid at the scale of micrometer. 
Solution to this equation, provided that the fluid motion obeys the no-slip boundary conditions on the walls, will reveal the detail form of the 
friction tensor. As there is no exact solution to the Stokes equation in an asymmetric channel, we will use approximate prescriptions for the  
friction coefficients. 
\begin{figure}
\centering
\resizebox{0.48\textwidth}{!}{%
  \includegraphics{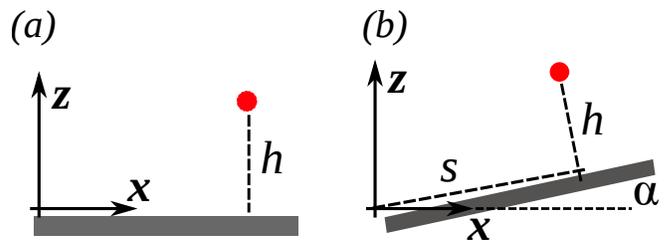}}
\caption{\small Left: A spherical particle with size $a$ located in a position with a distance $h$ above  a flat wall. Right: Same problem, but 
seen in a frame of reference rotated with an angle $\alpha$ with respect to the wall.}
\label{fig2}
 \end{figure}

Before considering the complicated geometry of our channel, we start by presenting the results  for friction coefficients of a  
particle moving in a semi infinite fluid environment  that is bounded by a single wall. 
Instead of friction tensor, it is more convenient to work with the mobility tensor ${\bf M}$ that is the inverse of friction tensor when expressed 
in the matrix 
notation: ${\bf M}={\bf G}^{-1}$. Fig.~\ref{fig2}(a), shows a spherical particle with size $a$ immersed in a semi infinite fluid that 
has  a distance $h$ from a wall. Theoretical analysis based on perturbation theory can give a series expansion for the mobility tensor  
in terms of small parameter $(a/h)$. In the matrix notation, the mobility tensor has the following components:
\begin{equation}
 {\bf M}= \left( \begin{array}{cc}
  M_{xx}&M_{xz}  \\
  M_{zx}& M_{zz} \\
\end{array}\right).
\end{equation}
Symmetry considerations do not allow  non-zero off-diagonal elements for the mobility tensor: 
$M_{xz}=M_{zx}=0$ and the other components are given by \cite{swan2007simulation,HydrodynamicCoupling}:
\begin{eqnarray}
M_{xx}&=&\mu\left(1-\frac{9}{16}\left(\frac{a}{h}\right)+\frac{1}{8}\left(\frac{a}{h}\right)^{3}-\frac{1}{16}\left(\frac{a}{h}\right)^{5}+ 
{\cal O}(\frac{a}{h})^7\right),\nonumber\\
M_{zz}&=&\mu\left(1-\frac{9}{8}\left(\frac{a}{h}\right)+\frac{1}{2}\left(\frac{a}{h}\right)^{3}-\frac{1}{8}\left(\frac{a}{h}\right)^{5}+ 
{\cal O}(\frac{a}{h})^7\right),\nonumber
\label{eq:verticaldiffution }
\end{eqnarray}
where $\mu=1/(6\pi\eta a)$, is the self mobility of a spherical particle with size $a$ moving in a fluid with viscosity $\eta$. 
Using the above components for the mobility tensor, one can write a perturbation series for the friction tensor as:
\begin{equation}
{\bf G}={\bf G}_0+\delta{\bf G},
\end{equation}
where ${\bf G}_0=6\pi\eta a{\bf I}$ is the friction tensor for a sphere moving in an infinite 
fluid and ${\bf I}$ is the unit tensor (unit matrix in matrix representation). Here all of the corrections due to the presence of wall are collected in  $\delta {\bf G}$. 
Direct calculations can give the explicit mathematical form of this correction term. 

Fig.~\ref{fig2}(b), shows the same problem as above, but in a reference frame that is rotated with an angle $\alpha$ with respect to 
the wall. It is an straight forward 
geometrical calculation, to express the mobility tensor in the rotated frame. 
In terms of the mobility tensor in the plate frame (frame in which the axis are parallel and perpendicular to the wall), the mobility 
tensor in the rotated frame can be given by:
\begin{equation}
{\bf M}={\cal R}^{T}(\alpha)\cdot{\bf M}\cdot{\cal R}(\alpha),
\end{equation}
where the dot symbol represents the matrix multiplication rule and ${\cal R}(\alpha)$ is the 
rotation matrix with an angle $\alpha$ about  axis ${\hat {\bf x}}\times {\hat {\bf z}}$. 

How a particle moving in medium confined with the walls of a complicated channel, will response to an external force? Exact solutions to this question 
occurs only in a certain type of symmetric geometries. A cylindrical channel with infinite length and also a 
rectangular channel with infinite length and infinite depth are examples with exact solutions \cite{exact}.
Apart from the above symmetric geometries, there is no analytic solution for the hydrodynamic friction problem in an asymmetric channel. 
Among from different approximate methods, the direct superposition of the corrections from different walls, 
is the simplest approximate scheme that can give the effects of walls in a channel \cite{happel,hydrosuperposition}. 
For a particle moving in a channel with confining walls, the friction tensor may be written as:
\begin{equation}
{\bf G}={\bf G}_0+\sum_i\delta{\bf G}^i,
\end{equation}
where $\delta{\bf G}^i$ is the correction due to the $i$'th wall of the channel.
Using this approximation we can study the dynamics of a Brownian particle moving in the channel.
Of course the validity of this superposition method is a challenging issue and it  will break near the walls and specially at the corners.  
At distances very near to the wall, the short range hard core interaction dominates the dynamics of the particle 
and we do not expect to see any sharp effect from the breakdown of our approximation for hydrodynamic interactions. 
Secondly, because of the geometrical constraints, we will see that the particle does not allow to reach the corners at all. These 
will ensure us that the superposition approximation will correctly account the long range hydrodynamic interactions with the wall of 
channel and we expect to have a picture that is at least qualitatively correct.  

\section{Numerical simulation}
We define the current density as the number of particles per unit length of the opening that exit from the left side of the channel in a unit time. 
Density of current for the particles, is a quantity that reflects how the motion of particles is rectified. 
In terms of the average velocity of the particles, we can write the current density as:
\begin{equation}
J=n\langle v_x\rangle,
\end{equation}
where $n$ is the density of particles, number of particles per unit area in our two dimensional problem. 
As we are not interested about the hydrodynamic interactions between the particles, we put  a single particle in our channel and study its dynamics. 
For this case the density is given by: $n=(L_xL_z(1+\frac{1}{2}\Delta\tan\theta))^{-1}$, with $\Delta=L_x/L_z$. Averaging over 
${\cal N}$ realizations of the system, we can obtain the average velocity as:
\begin{equation}
\langle v_x\rangle=\frac{1}{{\cal N}}\sum_{i=1}^{{\cal N}}\frac{x_i(t_s)-x_i(0)}{T},
\end{equation}
where $t_s$ is the time of observation (simulation time). 

To numerically solve the Langevin differential equation (Eq. \ref{langevin}), we need to make it non dimensional. 
For this purpose we can use $a$ as a scale for length and $\tau=a^2/(\mu k_BT)$ as a scale for time and 
make all variables non dimensional. 
After going to the dimensionless system, all of dynamical equations can be written in terms of dimensionless variables and some  
dimensionless numbers.
Denoting the non dimensional variables with an over bar, the non dimensional Langevin equation reads:
\begin{equation}
{\cal R}e{\bar {\bf r}}''={\bar {\bf G}}\cdot{\bar {\bf r}}'+{\bar f}_0\sin({\bar \omega}{\bar t}){\hat x}+{\bar {\Gamma}},
\end{equation}
where prime symbol denotes the derivative with respect to non dimensional time and  ${\cal R}e=m\mu/\tau$ is dimensionless 
Reynolds number and it measures how  the acceleration term in the dynamical equation of the particle is important  
with respect to a typical friction force. It is straightforward to see that at the scale of micrometer, the Reynolds number is very 
small: ${\cal R}e=10^{-3}$ . Non dimensional amplitude of the external force is given by 
${\bar f}_0=(\mu\tau/a)f_0$ and ${\bar {\bf G}}=\mu{\bf G}$. The noise satisfies the following correlation function:
\begin{equation}
\langle{\bar \Gamma}({\bar t}){\bar \Gamma}({\bar t}')\rangle=2{\bar {\bf G}}\delta({\bar t}-{\bar t}').
\end{equation}

Non dimensional current density can be written as ${\bar J}=J/J_0$, where $J_0=(\tau L_z(1+\frac{1}{2}\Delta\tan\theta))^{-1}$. 
To calculate the current density and average over realizations, we can use the periodic boundary condition and extract the 
average value of the current  from a  long time dynamics of a single particle.     
Please 
note that choosing the above  sort of non dimensional system, allows us to write the  average current density as:
\begin{equation}
{\bar J}=\frac{N_R-N_L}{N},
\end{equation}
where $N_R$ is the number of times  that the particle exits from the right opening and $N_L$ is the number of times that the particle 
enters to the channel from left side. Here $N=t_s/\tau$ and $t_s$ stands for the simulation time.

We use the method of reference \cite{ermak} to numerically integrate the equations of motion. To generate random numbers, we 
use the  Mersenne Twister pseudo-random number generator. 

Numerical parameters that  define our system  are as follows: we choose a spherical particle with size $a=1~\mu\mathrm{m}$ 
moving in a channel that is filled with water and it is at room 
temperature.   The viscosity of water is $\eta=10^{-3}\mathrm{Pa.sec.}$ and thermal energy is $k_BT=0.02~\mathrm{ev}$. 
These will result a characteristic  time scale that is 
$\tau=10~\mathrm{sec.}$. We choose a time step $\Delta t=0.005~\mathrm{sec.}$  that in dimensionless units   is about $\Delta \bar{t}=0.0005$. 
Diffusion times along the axis of channel and along a direction that is perpendicular to its axis are defined by $\tau_{D}^{x}=(L_x/a)^2\tau$ and $\tau_{D}^{z}=(L_z/a)^2\tau$  respectively. 
In a  channel with typical length $L_x=15 a$, we will see that $\tau_{D}^{x}=2000~\mathrm{sec}$. 
In the following section we will present the results of numerical calculations.

\begin{figure}
\centering
\resizebox{0.5\textwidth}{!}{%
\includegraphics{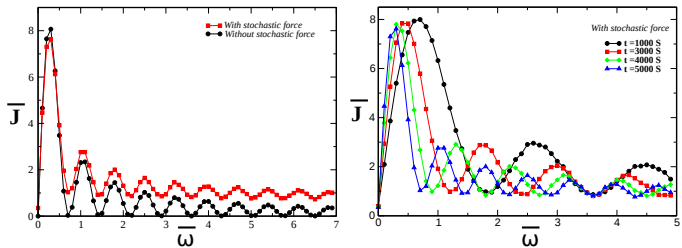}
}
\caption{Dimensionless  current density is plotted as a function of frequency of external force. Left: simulations have been performed for two cases where thermal 
force is present or absent. Simulation time is $t_s=5000~\mathrm{sec.}$ and for different simulation time is plotted.
Right: in the absence of thermal noise, the results are shown for different simulation times.
In all graphs $\bar{f}_0=10$ and  the geometrical variables are $a=1\mu\mathrm{m}$, $L_x=15a$, $L_z=5a$. 
}
\label{fig3}
\end{figure}

\section{Results}
Before investigating the hydrodynamic effects of the walls, we first turn off the hydrodynamic interaction 
and consider only the elastic collisions with the walls. To 
illustrate the  functionality of Brownian ratchet, we have plotted the average current density 
as a function of frequency of the external force in fig. \ref{fig3}(left). The results have been shown 
for two cases where the random noise is turned on or turned off. The simulation time is $t_s=5000~\mathrm{sec}$. 
One should note that, for this  small simulation time that we have chosen here, 
only the results at very large frequencies are acceptable. 
To obtain true physical results, we should keep in mind that the simulation time should be large in a way that 
 $\bar{t}_s\bar{\omega}\gg 1$.
As we expect, the existence of thermal noise is essential for the 
functionality of the Brownian ratchet. In the absence of thermal noise, no rectification is expected.
This is evident in the results, where the current at very large frequencies (the acceptable results) disappears for the case 
where thermal noise is absent. 
Here the diffusion time along the axis of channel is about 
$\tau_{D}^{x}=15^2\tau\sim 2000\mathrm{sec}$ and 
the period of external force for a special point on the graph with $\bar{\omega}=1$, is given by $T=2\pi/\omega=(2\pi/1)\times\tau\sim 60\mathrm{sec}$. 
The value of current at this frequency is $\bar{J}=2$ which 
corresponds to a current of particles that is about $J=2\times J_0\sim 10^5 m^{-1}\mathrm{sec.}^{-1}$.
%Minimum requirements to have a working ratchet are:
%\begin{equation}
%t_s\gg T,~~~t_s\gg\tau_{D}^{x},~~~t_s\gg\tau_{D}^{z}.
%\end{equation}
Here $t_s/\tau_{D}^{x}=2.5$, but as a result of external force, the particle has greater chance to travel many times along the  axes of channel. This would result an overall 
finite value   for both  $N_R$ and $N_L$.  

In fig. \ref{fig3}(right), we have studied the   system that is subjected to thermal noise, and investigated its response  by changing both the simulation time 
and frequency of external force. As one can see, by increasing the time of simulation, 
the height of all  non zero frequency peaks decreases. Increasing the simulations time will result a more smoother behavior for 
the current at large frequencies. Learning from the results of these graphs, we choose a proper simulation time and investigate the other physical properties of the Brownian ratchet in the 
following part.

\begin{figure}
\centering
%\resizebox{0.5\textwidth}{!}{%
\includegraphics[width=8cm]{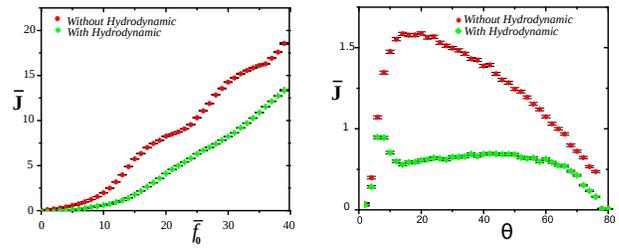}
\caption{Left: dimensionless current density as a function of the amplitude of external force is plotted for two cases where the hydrodynamic long range interactions 
with the walls are absent or present. Right: dimensionless current density as a function of the asymmetry angle $\theta$ is plotted for two cases where the hydrodynamic long range interactions 
with the walls are absent or present. 
Simulation time is $t_s= 10000\mathrm{sec}$, $\bar{f}_0=10$, $\bar{\omega}=100$, $L_x=15a$, $L_z=5a$ and $a=1\mu\mathrm{m}$.
}
\label{fig4}
\end{figure}

\begin{figure}
\centering
\resizebox{0.5\textwidth}{!}{%
\includegraphics{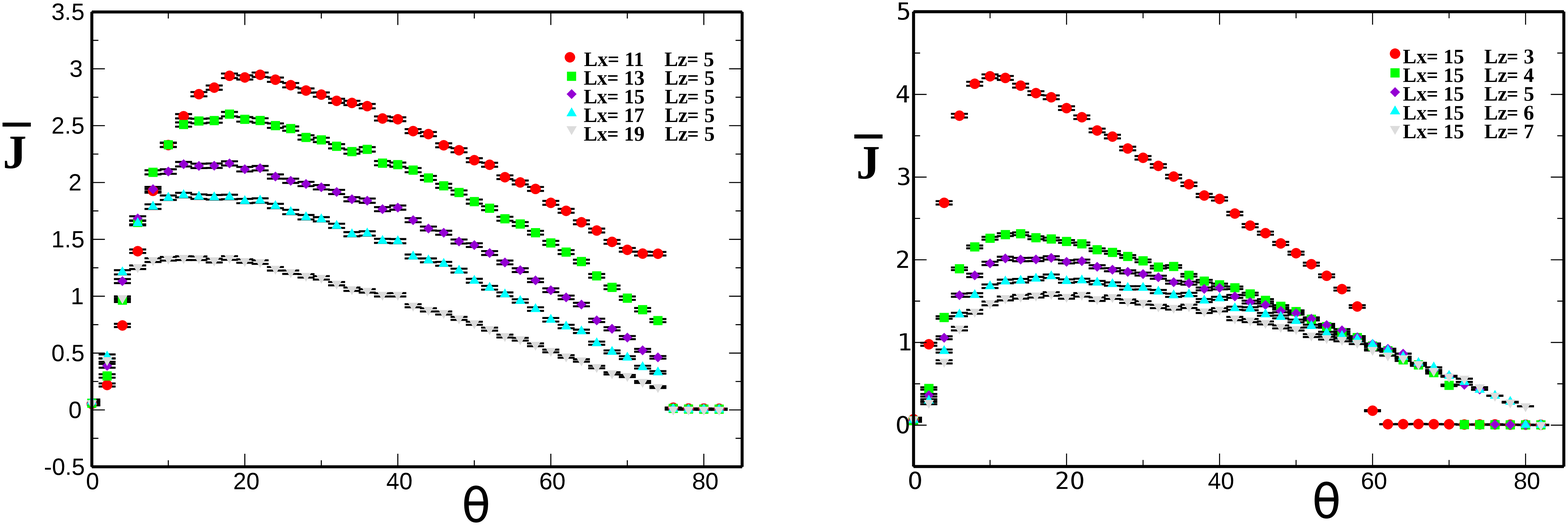}
}
\caption{Dimensionless current density as a function of the asymmetry angle $\theta$ for different values of $L_x$ and $L_z$ is investigated. 
Simulation time is $t_s= 10000\mathrm{sec.}$, $\bar{\omega}=100$ and $a=1\mu\mathrm{m}$.
}
\label{fig5}
\end{figure}
In fig. \ref{fig4}, we study how will the hydrodynamic interactions with the walls, modify the 
functionality of a Brownian ratchet. We have plotted the current density in terms of both amplitude of the 
externally applied force and also the asymmetry angle $\theta$. As one can see, in both cases the inclusion of 
long range hydrodynamic interactions with the walls, will significantly decrease the average current density. 
In terms of the amplitude of  external harmonic force,  the hydrodynamic interactions with the wall have more influences at large amplitudes. 
It is interesting that for a Brownian ratchet working at its efficient angle, the long range hydrodynamic interactions will deeply 
decrease the current.

This current reduction as a result of hydrodynamic interactions,   may  be  understood  in terms of the friction coefficients. 
As we have pointed out  
in section 2, for a particle moving near a single wall, the friction coefficient will be increased with a factor proportional to 
$(a/h)$. This means that for two particles, one moving in an infinite medium and the other moving near a wall, to achieve a constant velocity for both particles, 
higher force should be applied to the particle that is moving near wall. As the fluctuations are the source of rectified motion here, the 
increase in the friction coefficient will result a reduction in the current. This simplified picture that is inspired by a single wall confinement, can be applied qualitatively 
for a particle moving in a complex geometry of a channel. 

How do the geometrical parameters of this asymmetric  channel influence the functionality of Brownian ratchet?   
In fig. \ref{fig5}, we have investigated the influence of geometrical parameters of the system in 
current density  by changing both $L_x$ and $L_z$. The current density is plotted in terms of 
the asymmetry angle $\theta$. We  see that, for fixed $L_x$($L_z$),  higher 
current density can be achieved by choosing  smaller $L_z$($L_x$). In all cases, the maximum current density  corresponds to an 
angle $\theta\approx 20^{\circ}$. Interplay between the size of the particle and the available area in the corners, determines the overall 
behavior of the system.  Trapping time for the particle that is moving  in the corners of the channel will change the efficiency 
of a Brownian ratchet.  Dividing the area of the channel to two parts, one with a rectangular geometry with an area $L_x\times L_z$ and the second part with a 
triangle area, we see that  
less area for the rectangular part, corresponds to more current density.

In conclusion, we have considered the functionality of  a Brownian ratchet and investigate the role of hydrodynamic interactions on the efficiency of this system. 
To take into account the interactions with walls of the channel, we  proceeded with an approximate scheme for the mobility tensor of a colloidal particle moving 
near boundaries. We discussed about the limitations and validity of this approximation. As there is no exact solution for the mobility of a particle inside 
channel, to achieve more accurate results one can perform a simulation that includes a coupled equations of motion for colloidal particle and also the fluid 
particles. Although the results of such intense studies will help to obtain more accurate 
results, but we do not expect to see a large qualitative deviations with the results that we have obtained here.


\begin{thebibliography}{99}
\bibitem{smolo}
M. V. Smoluchovski, Phys. Z. {\bf 13}, 1069 (1912).

\bibitem{feynman}
R. Feynman, {\it Lectures on physics} (Addison Wesley Longman, 1997).

\bibitem{granular}
P. Eshuis, K. van der Weele, D. Lohse, and D. van der Meer, Phys. Rev. Lett. {\bf 104}, 248001 (2010).

\bibitem{ajdari}
F. Julicher, A. Ajdari, and J. Prost, 
Rev. Mod. Phys. {\bf 69}, 1269  (1997).

\bibitem{experimantal rocked ratchet}
N.S. Lin, T. W. Heitmann, K. Yu, B. L. T. Plourde, and V. R. Misko,  Phys. Rev. B \textbf{84}, 144511 (2011).

\bibitem{driftratchet}
C. Kettner, P. Reimann, P. Hanggi, and F. R. Muller,  Phys. Rev. E \textbf{61}, 312 (2000).


\bibitem{reimann2002brownian}
P. Reimann, Phys. Rep, \textbf{361}, 57-265 (2002).

\bibitem{particlespration}
D. Reguera, A. Luque, P. S. Burada,G. Schmid, J. M. Rubi\', and P. Hanggi,  Phys. Rev. Lett. \textbf{108}, 020604 (2012).

\bibitem{hanggi}
S. Martens, G. Schmid, A. V. Straube, L. Schimansky-Geier, and P. Hanggi,  Eur. Phys. J. Special Topics \textbf{222}, 2453 (2013).

\bibitem{martis}
S. Martens, I. M. Sokolov, and L. Schimansky-Geier, J. Chem. Phys   \textbf{136}, 111102 (2012).

\bibitem{review}
P. Hänggi and  F. Marchesoni, Rev. Mod. Phys. \textbf{81}  387 (2009).

\bibitem{holger}
A. Grimm and H. Stark, Soft Mater \textbf{7}, 3219 (2011).

\bibitem{pralle}
A. Pralle {\it et al.}, Appl. Phys. A {\bf 66}, S71 (1998).

\bibitem{squires}
E. R. Dufresne, T. M. Squires, M. P. Brenner, and D. G. Grier, Phys. Rev. Lett. {\bf 85}, 3317  (2000). 

\bibitem{roughwall}
S. H. Rad and A. Najafi, 
Phys. Rev. E {\bf 82}, 036305 (2010).


\bibitem{suspensionofconfined}
J.P. Hernandez-Ortiz, C.G. Stoltz, and M.D. Graham, Phys. Rev.  Lett. {\bf 95}, 204501 (2005).

\bibitem{3S2W}
A. Najafi, S. S. H. Raad, and R. Yousefi, 
Phys. Rev. E {\bf 88}, 045001 (2013). 


\bibitem{hnngiprl}
S. Martens, A. V. Straube, G. Schmid, L. Schimansky-Geier, P. Hanggi,  Phys. Rev. Lett. \textbf{110}, 010601 (2013).

\bibitem{swan2007simulation}
J. W. Swan and J. F. Brady,  Phys. Fluids \textbf{19} 113306 (2007).

\bibitem{HydrodynamicCoupling}
E. R. Dufresne, T. M. Squires, M. P. Brenner, and D. G. Grier, Phys. Rev. Lett. \textbf{85}, 3317 (2000).

\bibitem{exact}
N.~Liron and S.~Mochon, J. Eng. Math.  {\bf 10}, 287 (1976).


\bibitem{happel}
J. Happel and H. Brenner, {\it Low Reynolds Number Hydrodynamics},
(Noordhoff, Leyden, 1973).

\bibitem{hydrosuperposition}
B. Lin, J. Yu, and S. A. Rice, Phys. Rev. E {\bf 62}, 3909 (2000).

\bibitem{ermak}
D. L. Ermak and J. MacCammon, J. Chem. Phys \textbf{69}, 1352 (1978).


\end{thebibliography}
\end{document}